# Image-based Modelling of Organogenesis

Dagmar Iber[1,2,*], Zahra Karimaddini[1,2], Erkan Ünal[1,2,3]

**Affiliations:**

1 – Department of Biosystems, Science and Engineering (D-BSSE), ETH Zurich, Switzerland

2 - Swiss Institute of Bioinformatics (SIB), Mattenstraße 26, 4058 Basel, Switzerland

3 – Department of Biomedicine, University of Basel, Mattenstraße 28, 4058 Basel, Switzerland

* Corresponding Author: Dagmar Iber (dagmar.iber@bsse.ethz.ch)

**Summary Statements:**

- Methods for image-based modelling to explore Organogenesis
- A pipeline for the extraction of boundaries and displacement fields from 2D and 3D microscopy images
- Strategies for meshing and the solution of partial differential equations (PDEs) on the extracted (growing) domains

# ABSTRACT

One of the major challenges in biology concerns the integration of data across length and time scales into a consistent framework: how do macroscopic properties and functionalities arise from the molecular regulatory networks - and how can they change as a result of mutations? Morphogenesis provides an excellent model system to study how simple molecular networks robustly control complex processes on the macroscopic scale in spite of molecular noise, and how important functional variants can emerge from small genetic changes. Recent advancements in 3D imaging technologies, computer algorithms, and computer power now allow us to develop and analyse increasingly realistic models of biological control. Here we present our pipeline for image-based modeling that includes the segmentation of images, the determination of displacement fields, and the solution of systems of partial differential equations (PDEs) on the growing, embryonic domains. The development of suitable mathematical models, the data-based inference of parameter sets, and the evaluation of competing models are still challenging, and current approaches are discussed.

# INTRODUCTION

The process by which embryos self-organize into mature organisms with their particular shape and organs has fascinated scientists for a long time. Decades of experiments in developmental biology have led to the identification of many of the regulatory factors that control organogenesis and tissue development [1, 2]. Although these factors and their main regulatory interactions have been defined, the regulatory logic is often still elusive. A deeper understanding of the regulatory networks that govern developmental processes is important to enable the design and engineering of new functionalities and the repair of organs and tissues in regenerative medicine.

The size and morphology of organs and organisms are the result of the net effect of cell proliferation, differentiation, migration and apoptosis, and must be strictly controlled. Coordination of cell behavior within the tissue is achieved by cell-cell communication [3, 4] either between neighbouring cells via cell surface proteins (Notch-Delta, Cadherins etc) or gap junctions, or over longer distances via diffusible, secreted factors, called morphogens [5], or by cellular processes, referred to as cytonemes [6]. Mechanical feedbacks can further coordinate cell behavior within tissue [7-9].

Spatio-temporal gene expression and cellular events during morphogenesis are highly dynamic and involve complex interactions. Mathematical modeling has identified a number of candidate mechanisms for the control of such complex dynamic patterning processes [10, 11]. One of the oldest and most prominent ideas in developmental patterning is the Turing patterning mechanism. Alan Turing recognized that chemical substances, which he termed morphogens, can lead to the emergence of spatial patterns from noisy initial conditions by a diffusion-driven instability [5]. This mechanism has been highly attractive to explain biological patterning phenomena because deterministic, reliable symmetry breaks can be obtained with noisy initial conditions. A large number of patterning phenomena have been suggested to be controlled by a Turing mechanism, including those in the following references [12-16]. While all these models provide a good match with experimental observations, an experimental confirmation of the proposed Turing mechanisms is still outstanding because the kinetic rate constants have so far proved impossible to measure in the tissue. Pattern likeness itself is not a proof for the applicability of a Turing mechanism, and subsequent experiments in the *Drosophila* embryo have for instance ruled out the applicability of a Turing mechanisms – in spite of excellent pattern likeness [17]. More quantitative experimental data is therefore very much needed to test Turing models and alternative models, where proposed [13, 18].

The second important concept is the so-called French-flag model [19]. Assuming that a symmetry break has already taken place such that a particular morphogen is secreted only in some part of the tissue, this model stipulates that cells may determine their position relative to this source (and thus their cell fate) by reading out the concentration gradient that the morphogen will create across the tissue. Also here many open questions remain, in particular regarding the precision of the patterning mechanism [20-22], and its robustness to noise [23] and to variable environmental conditions, e.g. temperature, as well as to differences in the size of the embryonic domain [24-27].

Many further concepts have been analysed [10], and have been applied to various patterning phenomena, including the emergence of travelling waves [28], wave pinning [29], shuttling [30], chemotaxis [31, 32], and mechanical feedbacks [9, 33]. Further experimental testing and refinement of all these mathematical concepts and ideas will be important.

Most data in developmental biology is based on images, and recent developments in microscopy further enhance the opportunities to visualize developmental processes [34, 35]. Imaging enables the spatio-temporal visualization of regulatory aspects such as gene expression domains and protein localization, cell-based phenomena such as cell migration, division, and apoptosis [36], and the quantification of mechanical effects during morphogenesis [37-41]. Moreover, high-quality 3D imaging now provides detailed information on the complex architecture of tissues and organs [42, 43]. Most of these aspects have so far been analysed in isolation. Image-based modeling provides an opportunity to integrate the wide range of image-based spatio-temporal data with other biological datasets, e.g. qPCR, RNASeq, and proteomic data, into a consistent framework to arrive at a mechanistic understanding of morphogenesis [12, 14, 44-47]. Here, the gene expression domains indicate where the different model components are produced, while spatio-temporal information on protein distribution provide information regarding the spatio-temporal distribution of the model components. Moreover, cell-based data provides information on tissue growth and thus on advection and dilution terms (see below). In the following, we will explain the current approaches and methods for image-based modeling.

## SPATIO-TEMPORAL MODELS OF ORGANOGENESIS

Spatio-temporal models of regulatory networks can be formulated as a set of reaction-diffusion equations of the form

$$\frac{\partial c_i}{\partial t} = D\Delta c_i + R(c_k). \tag{1}$$

Here, $c_i$ denotes the concentration and $D_i$ the diffusion coefficient of component i. $\Delta$ represents the Laplace operator of the diffusion term, and $R(c_k)$ represents the reactions that alter the concentration of $c_i$. The spatio-temporal models must be solved on suitable domains, and boundary and initial conditions need to be specified. While initial values are typically not known, reasonable approximations can typically be made. Simulations may then need to be run for a certain time period to achieve a reasonable starting point for physiological simulations, e.g. a steady-state. Boundary conditions can be implemented either as a set number (Dirichlet boundary conditions), a flux (von-Neumann boundary condition), or a combination of the two. Sometimes, the domain of interest needs to be embedded into a larger domain to obtain realistic boundary effects for the (smaller) domain of interest [48]. As domains, idealized approximations to reality are typically employed [47]. Advances in microscopy and computer algorithms now, however, also permit an image-based approach, where computational domains are obtained from microscopy. In the following, we will present a pipeline for image-based modeling (Figure 1).

## A PIPELINE FOR IMAGE-BASED MODELLING OF ORGANOGENESIS

Before the computational work can be started, imaging data needs to be obtained of the tissue of interest. If different sub-structures are of interest, the tissue needs to be labeled accordingly. The sample preparation and staining methods of choice depends on the tissue, the substructure of interest, and the image acquisition technique and have their own challenges, which we will not discuss here. A large range of imaging techniques are available and are still being developed, as reviewed elsewhere [49]. The choice of imaging software is as important as the choice of the hardware. Thus, specialised software is required to control the functions of the optical microscopy tools and to collect and store the images that are acquired over time. A list of commercial, open-source and microscope companies' software packages have been reviewed elsewhere [50]. Once the imaging data has been obtained, the images have to be visualized and analyzed to extract meaningful quantitative data from the microscopy images. Algorithms have been developed to analyse and process images, and we refer the interested reader to the wide range of available textbooks and further texts on the topic, e.g. [51-53]. For the user, software packages are available, including 3-D Slicer, AMIRA, Arivis, BioImageXD, Icy, Drishti, Fijii, IMARIS, MorphoGraphX, Rhino, and Simpleware. For a review of the available software packages, bioimaging libraries, and toolkits with their applications we refer the reader to [50, 54]. In the following, we will provide an overview of the specific steps that need to be taken for image-based modeling, and we will describe the procedure in detail for the software package AMIRA (Box 1).

To use images in simulations, the following steps need to be performed: 1) image alignment and averaging, 2) image cropping and segmentation, and, 3) in case of growth processes, determination of the mapping between images at subsequent developmental stages. The domains can subsequently be meshed and used to solve mathematical models of the regulatory networks on the physiological domain. The finite element method (FEM) is the numerical method of choice to solve the PDE models on the physiological domains. A large number of FEM solvers are available, including Abaqus, ANSYS, and COMSOL Multiphysics as commercial packages and AMDiS, deal.ii, DUNE, FEniCS, FreeFEM, LifeV as open source academic packages. We will focus on the FEM solver COMSOL. In the following, we will discuss the different steps in detail.

### *Alignment & Averaging*

If multiple image recordings of the organ or tissue are available at a given stage, then the 3D images can be aligned and averaged. The alignment procedure is a computationally non-trivial problem and requires the use of suitable optimization algorithms, as described in textbooks and reviews [51, 55] (Box 1). Averaging is subsequently performed by averaging pixel intensities of corresponding pixels in multiple datasets of the same size and resolution. This helps to assess the variability between embryos and identifies common features. It also reduces the variability due to experimental handling. However, averaging of badly aligned datasets can result in loss of biologically relevant spatial information. It is therefore important to run the alignment algorithm several times, starting with different initial positions of the objects, which are to be aligned.

### *Image Segmentation*

The next step is to crop out the region of interest and to perform image segmentation. During image segmentation the digital image is partitioned into multiple subdomains, usually corresponding to anatomic features such as particular gene expression regions or different tissue layers (Figure 2). A wide range of algorithms is available for image segmentation, including thresholding, region-based segmentation, edge-based segmentation, level-sets, atlas-guided approaches, mean shift segmentation, and we refer the interested reader to textbooks and reviews on the topic [51-53]. Most segmentation algorithms are based on differences in pixel intensity and thus require a high signal-to-noise ratio in the images (see Box 1). Further manual work may be needed to complete domains.

*2D Virtual sections and isosurfaces of 3D images*

3D simulations are expensive and simulations are therefore often better first run on 2D sections (Figure 2). Following the alignment of the specimens in 3-D, 2-D optical sections can be extracted (see Box 1).

**SIMULATION OF NETWORK MODELS**

To simulate network models on the realistic geometries, the following steps need be performed: 1) generation of meshes, 2) import of meshes into an FEM solver, and 3) implementation of the PDE models and simulation of the models. The three steps will be discussed in the following.

*Mesh Generation & Import*

To numerically solve the partial derivative equations (PDEs) that describe the signaling networks of interest, surface and volume meshes of sufficient quality must be generated (Figure 2). A large variety of algorithms, including advancing front methods, grid, quadtree, and octree algorithms, as well as Delaunay refinement algorithms have been developed to create suitable meshes in 2D and 3D with different shapes, e.g. triangular, tetrahedral, quadrilateral and hexahedral, and we refer the interested reader to textbooks on the matter [56]. Image processing and FEM software packages (see above) as well as specialized meshing software, e.g. BioMesh3D, Gmsh, and MeshLab, can be used for mesh generation and subsequent improvement.

The quality of the mesh depends on the mesh size and the ratio of the sides of the mesh elements. The linear size of the mesh should be much smaller than any feature of interest in the computational solution, i.e. if the gradient length scale in the model is 50 $\mu$m then the linear size of the mesh should be at least several times lower than 50 $\mu$m. Additionally, the ratio of the length of the shortest side to the longest side should be 0.1 or more. To confirm the convergence of the simulation, the model must be solved on a series of refined meshes. When the geometries are extracted from experimental images, then the resolution is often so high that the resulting meshes are very fine. Too dense meshes increase the simulation time. The resolution must therefore be decreased. If the original resolution is very high, this may have to be done in multiple steps. While doing this, it is important to avoid intersections and inverted mesh elements.

***PDE Solvers***

To exchange meshes between the image processing software and the simulation software, suitable file formats need to be chosen (Box 1). Once the meshes have been imported to an FEM solver, the PDE models can be solved on the domains of interest. Solving computational models of organogenesis on physiological domains can be computationally expensive and computational complexity should therefore be minimized as much as possible. This can be achieved, in particular, by minimizing the number of mesh elements and by smoothing sharp transitions and domain boundaries as much as possible. Details on how to solve PDE models of organogenesis and some approaches to reduce simulation time in COMSOL have been described in [57-59].

## SIMULATIONS ON GROWING DOMAINS

During embryonic development, tissue morphogenesis and signaling are tightly coupled. It is therefore important to simulate both tissue morphogenesis and signaling simultaneously in *in silico* models of developmental processes. The development of mechanistic models of tissue growth is challenging and requires detailed knowledge of the gene regulatory network, mechanical properties of the tissue, and its response to physical and biochemical cues. If these are not available, 3D image data can be used to solve the models on realistically expanding domains. Confocal microscopy [14, 60] and light-sheet microscopy [36, 61] can be used for live imaging. If light-sheet microscopy is not available and the specimen is too large for confocal microscopy, a developmental series based on fixed samples can be acquired with optical projection tomography (OPT) [62].

***Determination of displacement fields***

To describe the growing domains, suitable displacement fields need to be determined from the data. In the ideal case, tissue development can be imaged continuously, and single cells can be tracked during live microscopy [63]. In that case, the displacement field can be determined as the displacement field of those cells. However, in most applications such data is not available. In the latter case, approximate displacement fields can be obtained by determining a mapping between the tissue boundaries of two stages that have been extracted at sequential time points during development. Ideally, the spacing between two time points should be rather small, at least small enough that the deformations are small. However, the minimal distance between two time points is often dictated by experimental limitations, in particular when biological processes cannot be imaged continuously and separate specimen at staged developmental time points have to be used. Even when cultured organs are imaged continuously, frame rates may be limiting. Accordingly, suitable algorithms need to be employed that can deal with the experimental time series.

While there are, in principle, an infinite number of possible mappings between such consecutive shapes, there are a number of important constraints. First of all, mappings should generally not intersect because this would create numerical problems and would generally also not be realistic because cells (in epithelia) will normally not swap positions with cells at other positions.

Moreover, given that the physiological displacement field is not known, it is sensible to create a displacement field that is both numerically favourable and biologically sensible. A range of mapping algorithms can be employed and which algorithm is most suitable depends on details of the geometries and their deformations (Algorithm 1) [64].

### *Algorithms for estimating displacement fields*

Before applying a mapping algorithm between the curve $C_1$ at time t and the curve $C_2$ at time t+Δt it is important to carry out three pre-processing steps: If the structure has grown substantially in between the two measurements, it is often helpful to first stretch the curves uniformly such that the two curves are similar (Figure 3A-D). Next, the same number of points, N, should be interpolated on both boundary curves such that the Euclidean distance is identical between all points on each boundary curve. Finally, the direction, in which the points on $C_1$ and $C_2$ are sampled, should be the same. After applying the above pre-processing steps, a map between the interpolated points can be established. Mappings based on the minimal distance between the two curves are easy to implement, but frequently fail (Figure 3A) [60, 64]. If $C_1$ and $C_2$ are open boundary curves, then the uniform mapping algorithm is more suitable, where the mapping is established between an equal number of equidistant points on the two curves (Figure 3E, Algorithm 2). When the growth of the domain is not uniform, then the displacement field is, however, biased towards the direction of maximal growth (Figure 3F). In case of closed boundaries, we recommend the use of normal mapping, where the mapping is based on vectors that are normal to the original curve $C_1$ and end on the target curve $C_2$ (Figure 3G, Algorithm 3). If the normal vectors intersect (Figure 3G), then either a reverse-normal mapping can be employed (Figure 3H), or the diffusion-mapping algorithm can be used instead (Figure 3I-L, Algorithm 4). In case of the diffusion mapping, the steady state diffusion equation is solved in between the curves $C_1$ and $C_2$ (Figure 3I), and the displacement field is based on the start and end points of streamlines (Figure 3J), which never cross. Depending on the shape of the domain the diffusion mapping is better calculated from $C_1$ to $C_2$ (Figure 3K) or in the reverse direction (Figure 3L). While the diffusion mapping prevents the crossing of the displacement vectors, we note that the diffusion mapping is computationally more expensive and the resulting displacement fields are more likely to result in mesh distortions. Diffusion mapping should therefore only be used when normal mapping or minimal distance mapping fail.

If a domain contains subdomains, then the boundaries that divide the subdomains intersect (Figure 4A). Intersections need to be avoided and the intersected curve should therefore be split into two curves at the intersection point (Algorithm 5) [65]. This step has to be done for all intersected boundaries at time t and at time t+Δt (Figure 4B,C). The displacement fields for all independent segments can then be determined as discussed above.

We note that Algorithm 1 only works well as long as the shapes are relatively close. In case of larger differences, we used the landmark-based Bookstein algorithm [66], as illustrated in Figure 5 [14]. Here we obtained lung buds at different developmental stages (Figure 5A), and determined the displacement field (Figure 5B) and thereby the growth fields (Figure 5C) between those stages. We then imported the mesh of the earlier stage lung bud into the FEM solver and solved signaling models on the imported shape (Figure 5D). In certain cases, the maxima of the predicted signaling fields coincided with the maxima of the embryonic growth fields (Figure 5E).

We note that while the landmark-based Bookstein algorithm [66] provided a straightforward method to obtain the displacement fields for the shapes, the usage of such a manual approach is very work intensive and dependent on the user and thus to a certain extent arbitrary. Moreover, for complex structures such as branched organs, the identification of mapping points can be difficult even when stereoscopic visualization technologies are available. We therefore recommend to adapt (semi-) automatic mapping algorithms as described above to the particular situation as much as possible.

### *Simulations on growing domains*

Once the displacement fields have been determined, they can be used to solve the PDE models on a realistically growing and deforming domain (Figure 6). To carry out the FEM-based simulations, the meshed geometries (Figure 6A) and displacement fields need to be imported into an FEM solver. On growing domains, advection and dilution effects need to be incorporated [67], such that Eq. (1) becomes on a growing domain

$$\frac{\partial c_i}{\partial t} + \nabla \cdot (\vec{u} \cdot c_i) = D\Delta c_i + R(c_k). \qquad (2)$$

Here $\vec{u}$ denotes the velocity field. To solve coupled PDE models on complex growing geometries, the Arbitrary Lagrangian Eulerian (ALE) method should be used. The ALE method is available in the commercial FEM solver

COMSOL, which we use to solve our PDE models of organogenesis on growing domains (Figure 6B) [15, 26, 48, 60, 68-70].

**PARAMETER ESTIMATION AND MODEL SELECTION FOR IMAGE-BASED MODELLING**

In the final step, the image-based data can be used to infer parameter sets for the models and to evaluate models and hypotheses. To evaluate the match between simulations and data, first a suitable metric needs to be defined to quantify the difference between model prediction and experimental data. Which metric is most suitable depends on the type of experimental data and the model [14, 71, 72]. Thus, only very little data is truly quantitative, such as the strength of the growth fields discussed above. A lot of data is purely qualitative, e.g. in situ hybridization data, and some data is semi-quantitative, e.g. fluorescent data. Given that the choice of a metric is to a certain extent arbitrary, but can affect the ranking of parameter sets and models, different metrics should ideally be evaluated [14, 71, 72]. For most models, several data sets need to be combined, and appropriate weighting functions need to be defined.

Once a metric and a weighting of the different datasets has been defined, an optimization algorithm has to be chosen to minimize the distance between the data and the simulation output. If a prior probability distribution for the parameter values is known, Bayesian optimization is the method of choice to incorporate this information into the optimization process. A wide range of global and local optimization algorithms have been developed, and we refer the interested reader to textbooks for details [73] and to previous reviews on parameter estimation for biological models [74-77]. If the models are sufficiently simple, then it is computationally feasible to first globally screen the parameter sets over the entire physiological range of the parameter values, and to subsequently carry out a local optimization procedure for the best parameter sets to further improve the fit between simulation output and data [14]. Based on the parameter screens, the posterior probability distributions for the parameters can be reconstructed and the relative likelihood of different models can be evaluated [78]. In case of multiple objectives (because of distinct datasets, genotypes etc.) in can be sensible to evaluate the Pareto optimality [71].

In case of more complex signaling models, a dense sampling of the parameter space is no longer possible. Interpolation algorithms such as Gaussian Process Upper Confidence Bound (GC-UCB) and Kriging methods can be used to guide the sampling process and thus reach the optimal parameter set with less sampling steps [79, 80]. Nonetheless it is sensible to keep model complexity to a minimum, and the field of model reduction is currently very active as reviewed elsewhere [81, 82]. Many biological models are modular and modules may be important on different time scales. In that case, parameter values can be estimated sequentially, as done recently in a tree-based approach [83].

**CONCLUSION**

Recent advances in imaging techniques, algorithms and computer power now permit the use of spatio-temporal imaging data for the generation and testing of increasingly realistic models of developmental processes. We have reviewed a pipeline for image-based modeling that comprises image acquisition, processing (segmentation, alignment, averaging), the determination of displacement fields, meshing of domains, transfer to an appropriate FEM solver, parameter inference and model selection. The computational approach reviewed here should enable a wider adoption of realistic spatio-temporal computational models by both biologists and computational biologists.

The field is rapidly moving forward and offers ample challenges and opportunities for computational scientists. Further developments are required, in particular in the areas of PDE solvers, cell-based models, and parameter inference. Thus, available PDE solvers were developed to tackle physical rather than biological problems and often cannot easily and efficiently be used to model organogenesis, especially on 3D growing domains. A number of software packages for cell-based tissue simulations already exist, but further developments are required to fully represent and analyse tissue mechanics and signaling in 3D [84]. Finally, current methods for parameter estimation cannot easily handle the complexity of biological models.

We expect that further developments in these three areas will allow to fully exploit the 3D dynamic imaging data that can now be obtained with light-sheet microscopy for a mechanistic understanding of developmental processes.

**Acknowledgements**

The authors thank past and present members of the CoBi group for discussions and acknowledge funding from a SystemsX iPhD grant and the SystemsX NeurostemX RTD.

# BOX 1: SOFTWARE DETAILS FOR IMAGE PROCESSING, MESH GENERATION & PARAMETER OPTIMIZATION

In the following, we provide details on AMIRA-based image processing and mesh generation as well as mesh import, PDE model simulation, and parameter optimization in the FEM solver COMSOL.

## 1. *Alignment & Averaging*

To align images a number of iterative hierarchical optimization algorithms, e.g. QuasiNewton, are available in the software package AMIRA, as well as similarity measures, e.g. Euclidean distance. Averaging is subsequently performed by averaging pixel intensities of corresponding pixels in multiple datasets of the same size and resolution.

## 2. Image Segmentation

In AMIRA, a variety of algorithms are available for image segmentation, including a thresholding algorithm, which is suitable for most 3-D images, a watershed algorithm for the segmentation of more complex structures, and a top-hat algorithm for the detection of dark or white regions [52, 85]. Given the dependency of these algorithms on differences in pixel intensity and thus on a sufficient signal-to-noise ratio in the images, further manual work may be needed to complete domains.

## 3. 2D Virtual sections and isosurfaces of 3D images

In AMIRA, the SurfaceCut module is available for this purpose. The boundaries of the 2-D sections can be extracted in AMIRA using the Intersect module. Once saved in ASCII format, the section boundaries can be further processed in MATLAB to smooth and fine-tune the subdomains of the boundaries.

### *4. AMIRA Mesh Import into the FEM solver COMSOL*

AMIRA meshes can be saved in the I-DEAS universal data format, which can be read by the Gmsh software and saved as a Nastran Bulk data file. The file extension of the Nastran Bulk data file needs to be changed from UNV to DAT to make it compatible with the simulation software COMSOL Multiphysics. In the COMSOL mesh module, the mesh can then be imported.

### *5. Parameter Optimization in COMSOL*

COMSOL provides a range of optimization solvers, including gradient-based ones, e.g. Levenberg-Marquardt, Method of Moving Asymptotes (MMA), Sparse Nonlinear Optimizer (SNOP), and derivative free ones, e.g. constraint optimization by linear approximation (COBYLA), and bound optimization by quadratic approximation (BOBYQA).

**BOX 2: MAPPING ALGORITHMS**

- Uniform Mapping (Algorithm 2)

    This algorithm maps each point on $C_1$ to a point on $C_2$ such that the order of points on $C_1$ and $C_2$ is conserved, assuming that both boundary curves contain the same number of equidistant points.

- Normal Mapping (Algorithm 3)

    The normal vector at every point on the curve $C_1$ is computed. Then, each point on $C_1$ is mapped to the closest point on C2, where the normal vector and C2 intersect.

- Diffusion Mapping (Algorithm 4)

    This algorithm solves the steady state diffusion equation ($\Delta c=0$) with Dirichlet boundary conditions (c=1 on $C_1$ and c=0 on $C_2$) in the area between the two boundaries. The mapping vectors are calculated as the difference between the start and the end point of the streamlines. The streamlines can be estimated using the COMSOL Multiphsyics particle tracking module.

- Intersecting Curves (Algorithm 5)

    In case of two boundary curves $B_1 = \{P_{11}, \ldots, P_{n1}\}$ and $B_2 = \{P, \ldots, P_{m2}\}$ that intersect at point P, $B_1$ should be split into two segments $B_{1,segment1} = \{P_{11}, \ldots, P\}$ and $B_{1,segment2} = \{P, \ldots, P_{n1}\}$, while $B_2$ remains the same. This step has to be done at time t and at time t+$\Delta$t. To determine the displacement fields, uniform mapping (Algorithm 2) should subsequently be applied to each open segment.

# FIGURES

## Figure 1: A Pipeline for Image-based Modelling of Organogenesis

The figure summarizes the steps for image-based modeling. For details see the main text.

## Figure 2: Extraction of physiological domains for FEM modelling

After acquisition of the 3D image of the organ of interest (here an embryonic lung), the structure of interest is identified (here the tip of a lung bud) and cropped out for further processing. Based on markers, sub-structures of interest can be segmented. Finally, surfaces can be extracted. In a separate step, 2-D surface cuts can be generated to extract 2D planes and their boundaries. Finally, the domain can be meshed.

## Figure 3: Determination of Displacement Fields

(A-D) Transformation prior to the mapping can improve the resulting displacement field. (A) The displacement field (red) was computed between $C_1$ (green) and $C_2$ (blue) using the minimal distance algorithm. (B) $C_1$ is scaled to be more similar to $C_2$; the new curve (red, dashed line) is called $C_{1,s}$. (C) The displacement field (red) is computed from the transformed curve $C_{1,s}$ (green) onto $C_2$. (D) The starting points of the vectors on $C_{1,s}$ where finally transformed back onto the original curve $C_1$. Comparison of the displacement fields (red) in panels A and D shows that, as a result of the transformation, the number of regions on $C_2$, onto which no points were mapped, has decreased and the displacement field quality has therefore increased.

(E-F) Uniform mapping. (E) The displacement field of open curves is determined by interpolating both curves with equidistant points and by mapping the points consecutively onto each other. (F) When the growth of the domain is not uniform, the displacement field is biased towards the direction of maximal growth.

(G-H) Normal mapping. (G) If $C_1$ encloses a non-convex domain and both curves are not very close to each other in the concave regions of $C_1$, the displacement field vectors generated by normal mapping cross each other. (H) The problem can be resolved by using reverse normal mapping, i.e. by mapping $C_2$ onto $C_1$.

(I-L) Diffusion mapping. (I) The steady-state solution of the diffusion equation. (J) The streamlines, as obtained by particle tracking from $C_1$ to $C_2$. (K) The displacement field vectors are obtained by connecting the start and the end points of the streamlines. (L) Reverse diffusion mapping, i.e. $C_2$ is mapped onto $C_1$. This yields better displacement fields when $C_2$ is much larger than $C_1$.

The figure reproduces panels from [64].

**Figure 4: Displacement Fields for Domains with Internal Boundaries**

(A) A domain with three subdomains. The boundary curves (black, red, green) intersect at different points. (B) $C_1$ (green) and $C_2$ (red) intersect at point P. (C) To map the intersection point P (blue point) at time t to the intersection point at t+1, the curves are divided into segments, such that the point P becomes the start/end point of the intersecting curves. The figure reproduces panels from [65].

**Figure 5: Image-based Modelling with embryonic data**

(A) 3D lung bud shapes at two different stages. (B) A displacement field between the two stages shown in panel A. (C) The displacement vectors with out-ward pointing vectors correspond to the growth field. (D) The solution of a ligand-receptor based Turing model on the first stage of the segmented lung in panel A. (E) The predicted distribution of the ligand-receptor signaling pattern (solid colours) coincides with the embryonic growth field (arrows). The Figure reproduces panels from [14].

**Figure 6: Simulations on growing domains**

(A) The meshed domain of an ureteric bud. (B) The solution of a computational model (solid colours) and the estimated growth fields (coloured arrows) of a growing ureteric bud. The figure reproduces panels from [64].

**Figure 1**

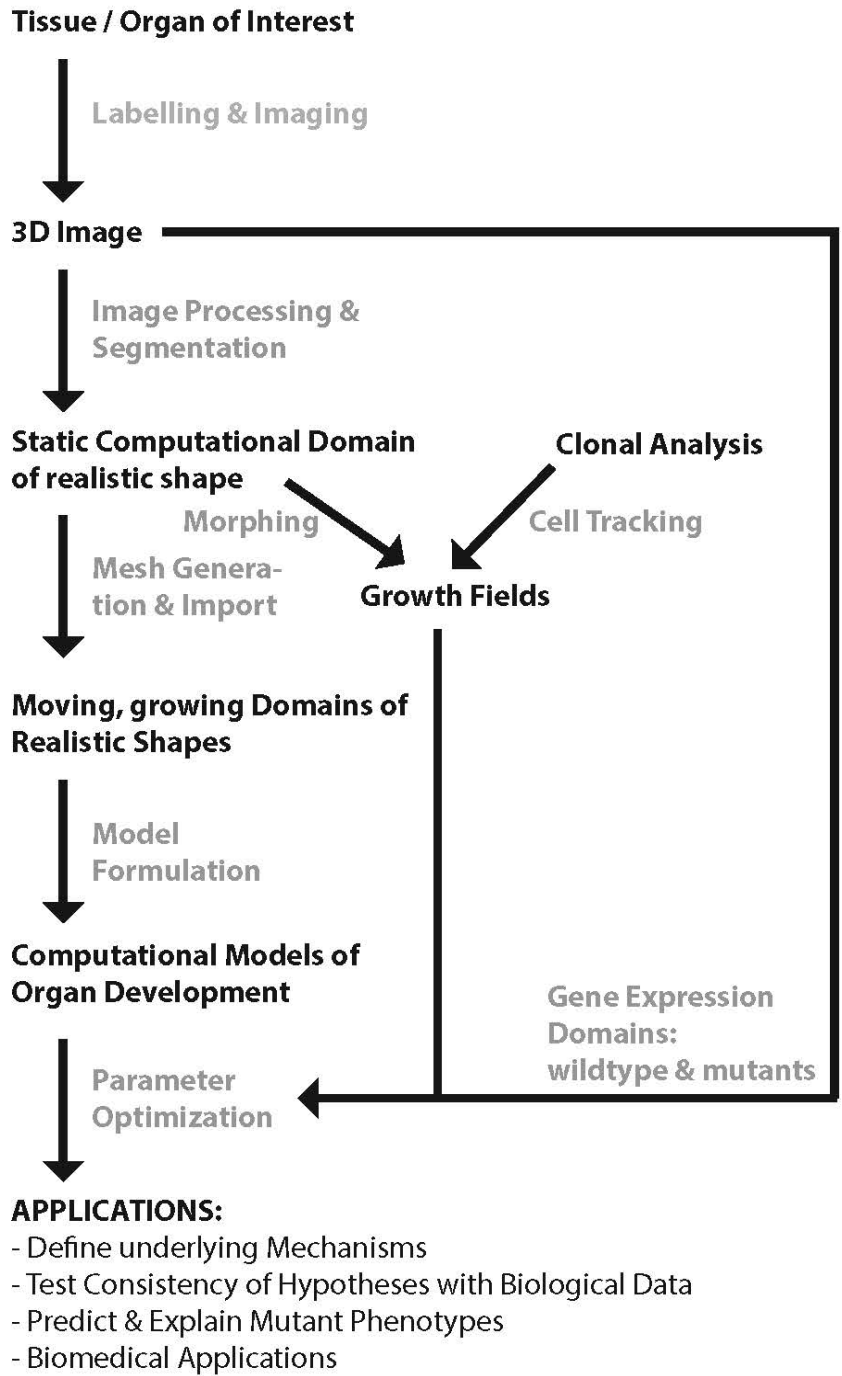

**Figure 2**

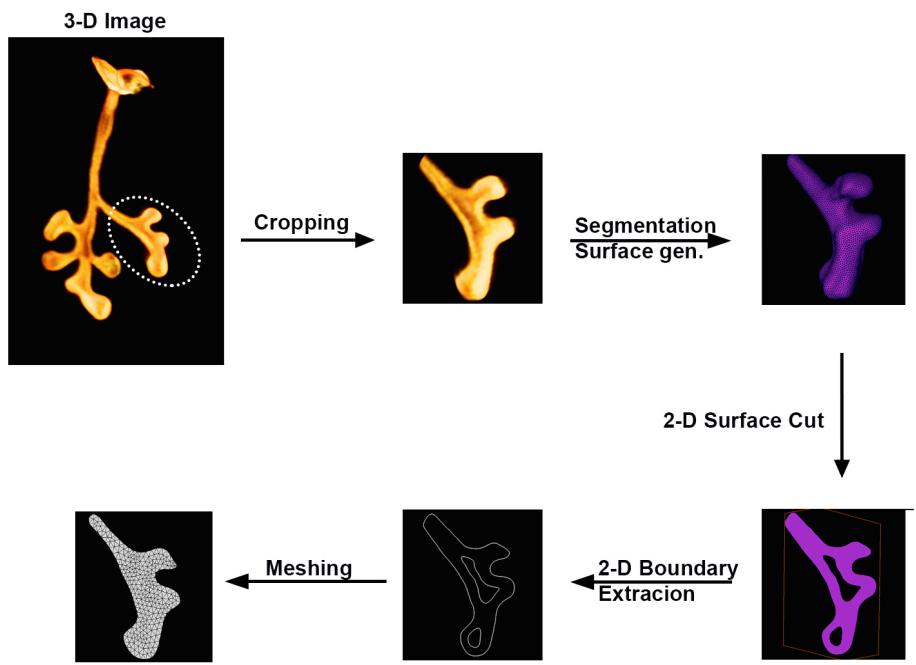

Figure 3

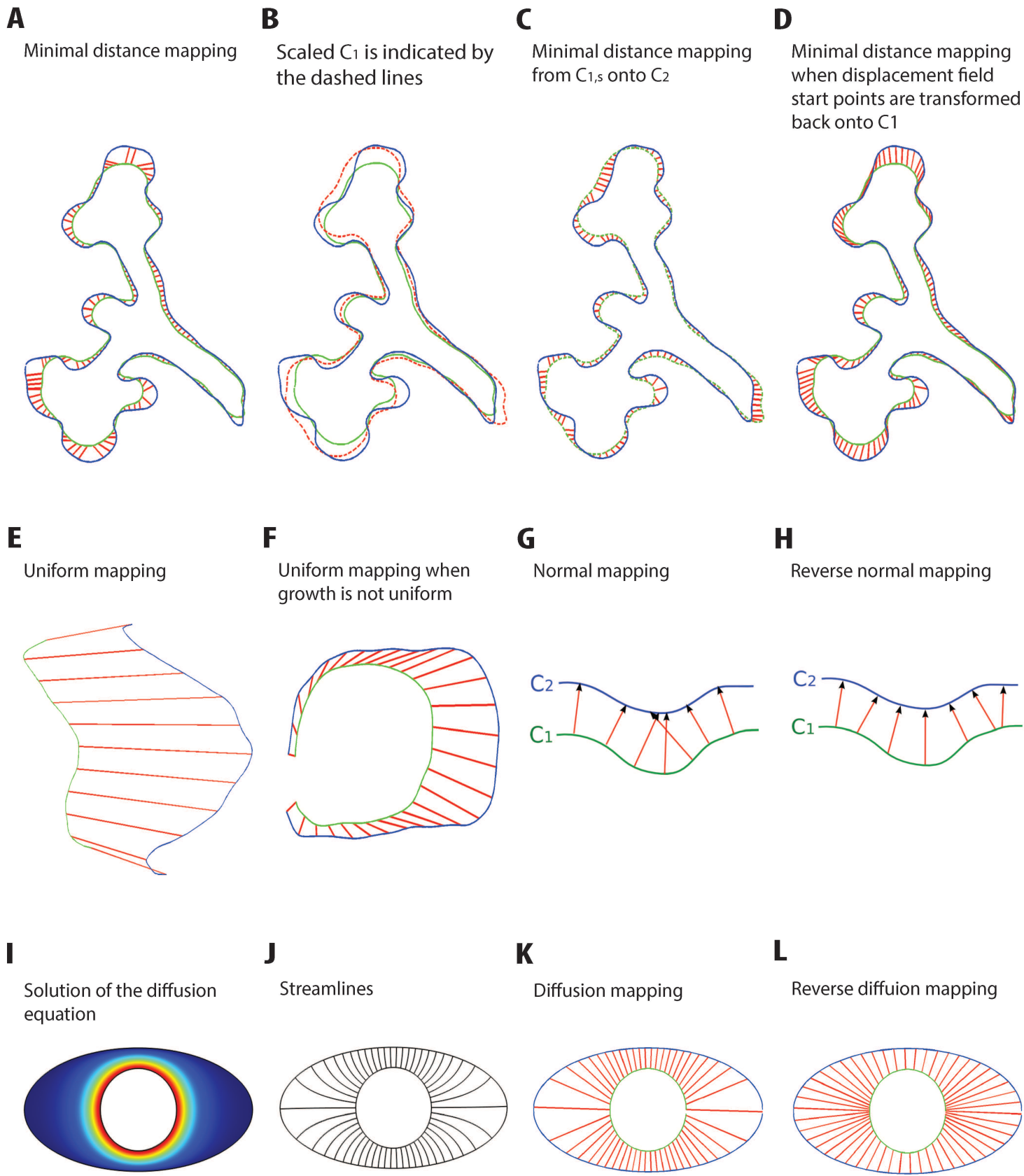

Figure 4

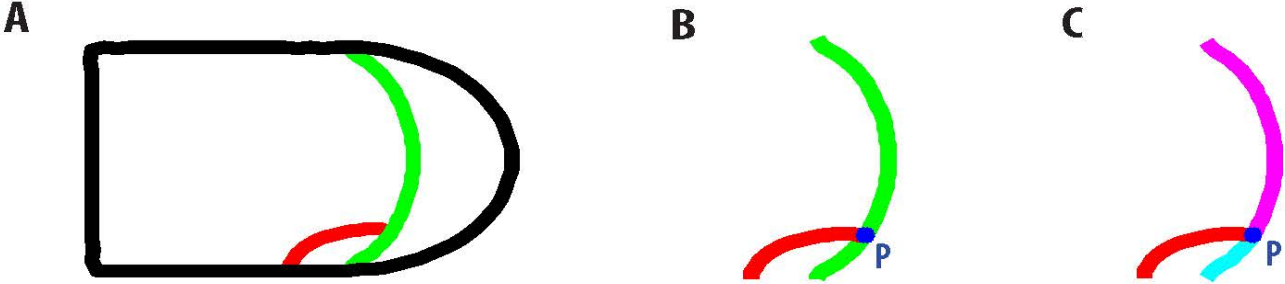

**Figure 5**

A

Stage 1   Stage 2

B   C

shrinkage   growth

D   E

**Figure 6**

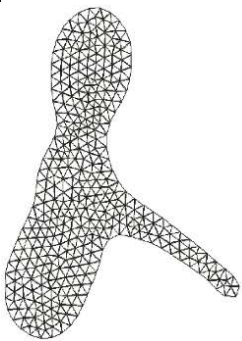 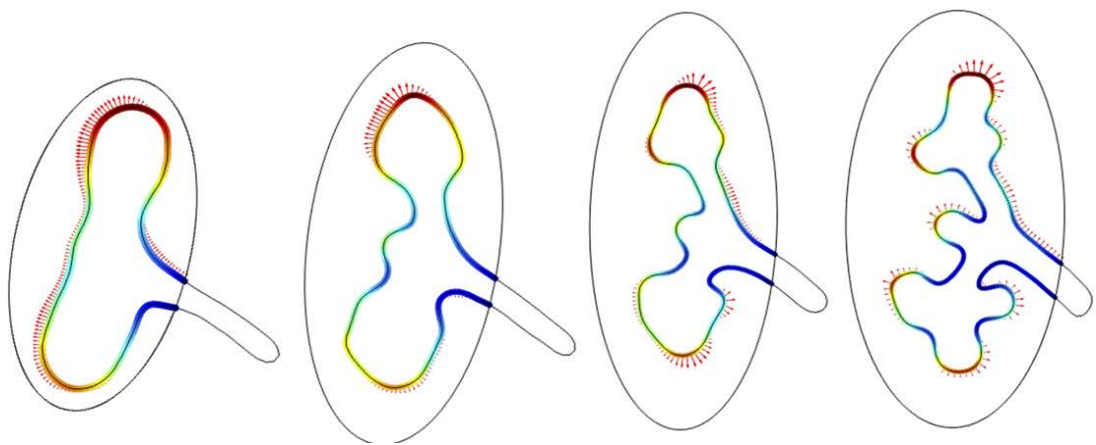